# Simulation of Particle-Material Interactions

*N.V. Mokhov*[1]
[1]Fermilab, Batavia, IL 60510, USA

**Abstract**
This paper gives an overview of the particle transport theory essentials, the basics of particle-material interaction simulation, physical quantities needed to simulate particle transport and interactions in materials, Monte Carlo simulation flow, response of additive detectors, statistical weights and other techniques to minimize statistical errors. Effects in materials under irradiation, materials response related to component lifetime and performance are considered with a focus on high-energy and high-power accelerator applications. Implementation of simulation of particle-material interactions in the modern Monte Carlo codes along with the code's main features and results of recent benchmarking are described.

**Keywords**
Particle transport and interaction with materials simulation; Monte Carlo method techniques; modern Monte Carlo codes features and performance.

## 1 Introduction

The consequences of controlled and uncontrolled impacts of high-intensity and/or high-power and/or high-energy beams on components of accelerators, beamlines, target stations, beam collimators, absorbers, detectors, shielding, and environment can range from minor to catastrophic. Strong, weak, electromagnetic and even gravitational forces (neutron oscillation and neutron TOF experiments) govern high-energy beam interactions with complex components in the presence of electromagnetic fields. Therefore, simulations are only possible with a few well-established Monte Carlo codes (no analytic or simplified approaches are used these days). Predictive power and reliability of particle transport simulation tools and physics models in the multi-TeV region should be well-understood and justified to allow for viable designs of future colliders with minimal risk and a reasonable safety margin.

Physics of interactions of fast particles with matter is described in detail in literature (e.g., in Refs. [1-8]). Electromagnetic interactions, decays of unstable particles and strong inelastic and elastic nuclear interactions all affect the passage of high-energy particles through matter. At high energies, the characteristic feature of the phenomenon is creation of hadronic cascades and electromagnetic showers (EMS) in matter due to multi-particle production in electromagnetic and strong nuclear interactions. Because of consecutive multiplication, the interaction avalanche rapidly accrues, then passes the maximum, and then dies as a result of energy dissipation between the cascade particles as well as ionization energy loss. Energetic particles are concentrated around the projectile axis forming the shower core. Neutral particles (mainly neutrons) and photons dominate with a cascade development when energy drops below a few hundred MeV. Low-energy neutrons coupled to photons propagate large distances in matter around cascade core, both longitudinally and transversely, until they dissipate their energy in a region of a fraction of an electronvolt. Muons - created predominantly in pion and kaon decays during the cascade development – can travel hundreds and thousands of meters in matter along the cascade axis. Neutrinos – usual muon partners in such decays – propagate even farther, hundreds and thousands of kilometers, until they exit the Earth's surface.

This paper is divided into two main sections. The first section gives a brief overview of the particle transport theory essentials. The second section describes implementation of simulation of particle-material interactions in the modern Monte Carlo codes along with the code's main features and results of recent benchmarking.

## 2 Basics of particle-material interaction simulation

### 2.1 Particle transport theory essentials

A fundamental quantity $N_i(\vec{r}, \vec{\Omega}, E, t)$ is differential in space, angle, energy and time density of particles of *i*-type in a unit phase volume near a phase point $\mathbf{x} = (\vec{r}, \vec{\Omega}, E, t)$ [3]. Below we use $N = N_i$. Let's consider elemental area $dS$ centered at a point $\vec{r}$ and having normal vector $\vec{n}$. For a time interval $dt$, the area can be crossed by particles contained in the volume $dV = |\vec{v}\,\vec{n}|\,dS\,dt = |\vec{\Omega}\,\vec{n}|\,v\,dS\,dt$, where $\vec{v} = v\vec{\Omega}$ is a velocity of a particle corresponding to its energy $E$. The number of such particles is

$$N(\mathbf{x}, t)\,dV = \vec{\Omega}\,\vec{n}\,\Phi(\vec{r}, \vec{\Omega}, E, t)\,dS\,dt,$$

where function $\Phi(\vec{r}, \vec{\Omega}, E, t) = v\,N(\vec{r}, \vec{\Omega}, E, t)$ is *differential flux density* for particles of *i*-type. Again, here we use here $\Phi = \Phi_i$. Many other characteristics used in particle transport theory can be derived from $\Phi(\mathbf{x})$. The following functionals of $\Phi(\mathbf{x})$ are frequently used:

$\Phi(\vec{r}, \vec{\Omega}, t) = \int_0^\infty \Phi(\vec{r}, \vec{\Omega}, E, t)dE$ is spatial-angular flux density

$\Phi(\vec{r}, E, t) = \int_{4\pi} \Phi(\vec{r}, \vec{\Omega}, E, t)d\vec{\Omega}$ is spatial-energy flux density

$\Phi(\vec{r}, t) = \int_0^\infty \Phi(\vec{r}, E, t)dE = \int_{4\pi} \Phi(\vec{r}, \vec{\Omega}, t)d\vec{\Omega}$ is flux density

$\Phi(\vec{r}) = \int_0^\infty \Phi(\vec{r}, t)dt$ is particle fluence

$\Phi(t) = \int \Phi(\vec{r}, t)d\vec{r}$ is particle flux

$E\Phi(\vec{r}, \vec{\Omega}, E, t)$ is differential energy flux density

$\vec{\Omega}\Phi(\vec{r}, \vec{\Omega}, E, t)$ is differential current density

$S(\vec{r}, \vec{\Omega}, E, t) = \Sigma(\vec{r}, E)\,\Phi(\vec{r}, \vec{\Omega}, E, t)$ is differential collision (absorption, scattering) density, where $\Sigma(\vec{r}, E)$ is collision (absorption, scattering) macroscopic cross-section, $\Sigma(\vec{r}, E) = \frac{1}{\lambda(\vec{r}, E)}$ where $\lambda(\vec{r}, E)$ is particle mean free-path

$S(\vec{r})$ is often called particle star density or density of inelastic nuclear interactions, if $\Sigma(\vec{r})$ is macroscopic absorption x-section.

$\Phi(\vec{r}, \vec{\Omega}, E, t)$ obeys the system of *Boltzmann equations* which are derived from a balance of particles of *i*-type (index *i* is skipped again) incoming to and leaving from a unit phase space along with creation and absorption:

$$\frac{1}{v}\frac{\partial}{\partial t}\Phi(\vec{r}, \vec{\Omega}, E, t) + \vec{\Omega}\,\vec{\nabla}\Phi(\vec{r}, \vec{\Omega}, E, t) + \Sigma(\vec{r}, E)\Phi(\vec{r}, \vec{\Omega}, E, t)$$
$$= \sum_j \int d\Omega' \int dE'\,\Sigma(\vec{r}, \vec{\Omega'} \to \vec{\Omega}, E' \to E)\Phi_j(\vec{r}, \vec{\Omega}, E, t) + G(\vec{r}, \vec{\Omega}, E, t) \qquad (1)$$

where $\Sigma(\vec{r}, E)$ is a *total macroscopic cross-section* for the given particle type (*i*); in addition to nuclear interaction, it can include the decay one for unstable particles $\Sigma_D(E) = \frac{1}{\lambda_D} = m/(c\tau p)$ with mass $m$ and lifetime $\tau$, as well as ionization loss one for charged particles $-\frac{\partial}{\partial E}\frac{dE}{dx}(\vec{r},E)/\rho(\vec{r})$; $\Sigma(\vec{r},\Omega' \to \Omega, E' \to E)$ is a *double-differential cross-section* which defines creation of particles of *i*-type in the end state $\vec{\Omega}, E$ from the initial state $\vec{\Omega'}, E'$ by all types *j* of particles under consideration; $G(\vec{r}, \vec{\Omega}, E, t)$ is the density of the external sources.

## 2.2 Monte Carlo methods in particle transport in matter

### 2.2.1 Deterministic and Monte Carlo methods

The Bolzmann equations were successfully used for decades in reactor applications for neutrons and photons in 1D, 2D and even 3D cases applying *deterministic transport methods* for the average particle behavior (the most common of which was the discrete ordinate method). Starting early 60's, several attempts have been undertaken to solve the system of transport equations at high energies for – as called at that time - "nucleon-meson cascades" [9-11]. Special forms of the $\Sigma(\vec{r}, \Omega' \to \Omega, E' \to E)$ double-differential cross-sections and a number of other simplifications allowed even analytical solutions of the system but only in the 1D case. It has become clear that the general case of the hadronic-electromagnetic cascades with all the known elementary particles and products of nuclear reactions can only be solved using Monte Carlo methods. The general case includes all the correlations at the interaction vertices over the energy range of $10^{14}$ decades, with cascades developed in complex 3D geometry of accelerator facilities, experimental setups and their detectors, with magnetic fields etc.

Nowadays, the *Monte Carlo method (MCM)* [3, 12-14] is the principal, if not only, method in particle transport in accelerator applications. In its simplest and, at the same time, most dependable and common form – *direct mathematical modelling* – it involves numerical simulation of the interactions and propagation of particles in matter. In this approach, all the physics processes are modelled as if these take place in the real world, in realistic geometry and fields of accelerators and experimental setups. The use of various modifications of MCM, the so-called *variance reduction techniques*, makes it possible to greatly simplify the solution of the problem in certain cases, with a very high accuracy reached in a phase space volume of interest.

### 2.2.2 Sampling random quantities

The random continuous quantity $\xi$ is determined in the interval $(a, b)$ by the function $p(x)$, which is called the *probability density*. The probability that $\xi$ will be in the interval $(a, x)$ is

$$P(a < \xi < x) = \int_a^x p(x')dx', \text{ where } p(x) > 0 \text{ and } \int_a^b p(x)dx = 1. \qquad (2)$$

The expectation value of $\xi$ (its *mean value*) is $\boldsymbol{M}\xi = \int_a^b xp(x)dx$. For random continuous function $f(x)$, $\boldsymbol{M}f(\xi) = \int_a^b f(x)p(x)dx$. With a generator of random numbers $\gamma$, which are uniformly distributed in the interval $(0,1)$, the use of the equation $\gamma = P(x) = \int_a^x p(x')dx'$ allows us to randomly choose the values $x = P^{-1}(\gamma)$. This the *inverse-function method* used when the integral can be expressed in terms of the elementary functions. Otherwise, the *Neumann (or rejection) method* can be used: the density function of $\xi$ ($a < \xi < b$) is redefined as follows $p^*(x) = \frac{p(x)}{Max[p(x)]}$. Choose two random numbers, $\gamma_1$ and $\gamma_2$, and calculate $x' = a + \gamma_1(b - a)$. If $\gamma_2 < p^*(x')$, $\xi = x'$. Otherwise, the $(\gamma_1, \gamma_2)$ pair is discarded, a new pair is chosen, and the procedure is repeated. This is how the MCM was used for the first time for calculation of the area (volume) of an arbitrary shape.

## 2.2.3 Nuclear interaction cross-section and double-differential cross-section

The first most important quantity to solve Eq. (1) or simulate particle transport by Monte Carlo method is a macroscopic cross-section for a given interaction type of a particle of *i*-type with a chosen target nucleus with atomic mass $A$

$$\Sigma(i, A, E) = \frac{\sigma(i,A,E) N_A \, 10^{-2} \, \rho}{A}, \text{cm}^{-1}, \quad (3)$$

where $\sigma$ is a microscopic x-section (mb) for a given reaction, $N_A$ is the Avogadro number (=6.022× $10^{23}$ mol$^{-1}$), $\rho$ is material density (g/cm$^3$). Figure 1 shows the total and elastic microscopic cross-sections for the $\pi^+ p$ interaction *vs* square root of the total energy of the reaction $s = m_1^2 + m_2^2 + 2E_{1lab} m_2$. The elementary cross-sections exhibit strong momentum dependence below 3 GeV/c, are almost flat up to about 200 GeV/c and grow logarithmically with momentum above 200 GeV/c. Nuclear interaction cross-sections show some momentum dependence below 3 GeV/c and are practically momentum-independent at higher momenta growing with atomic mass of a target nucleus (see Fig. 2).

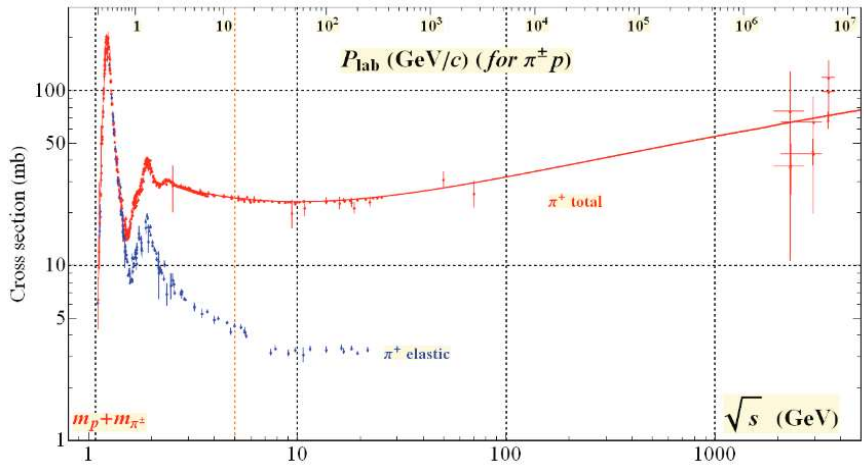

**Fig. 1:** Total and elastic $\pi^+ p$ interaction cross-sections as a function of the square root of total energy and pion momentum. Points are experimental data; lines are adopted parameterizations [15].

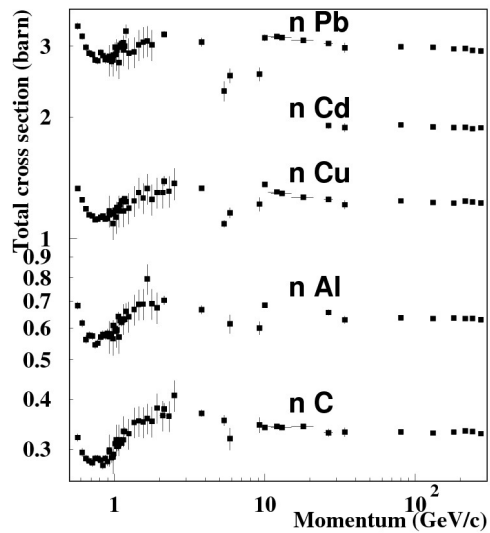

**Fig. 2:** Total cross-sections for neutron interactions on different nuclei as a function of neutron momentum [16].

When neutron energy falls below about 20 MeV, the physics of neutron interactions and transport are governed by those of the reactor domain. These are dominated by elastic and inelastic scattering, capture of low-energy neutrons with emission of photons, and fission on high-Z nuclei. These processes are correspondingly labeled in Fig. 3 for neutron interactions on $^3$He, $^{27}$Al, $^{56}$Fe, and $^{238}$U. One can see a characteristic resonant structure for neutron energy above 10 eV for $^{238}$U and 10 keV for $^{27}$Al up to about 10 MeV, quite different for different isotopes of the same element. Such a structure forces development and use in simulations of very detailed many-Gigabyte evaluated data libraries, (e.g., ENDF/B-VIII [17]), for macroscopic cross-sections and differential cross-sections.

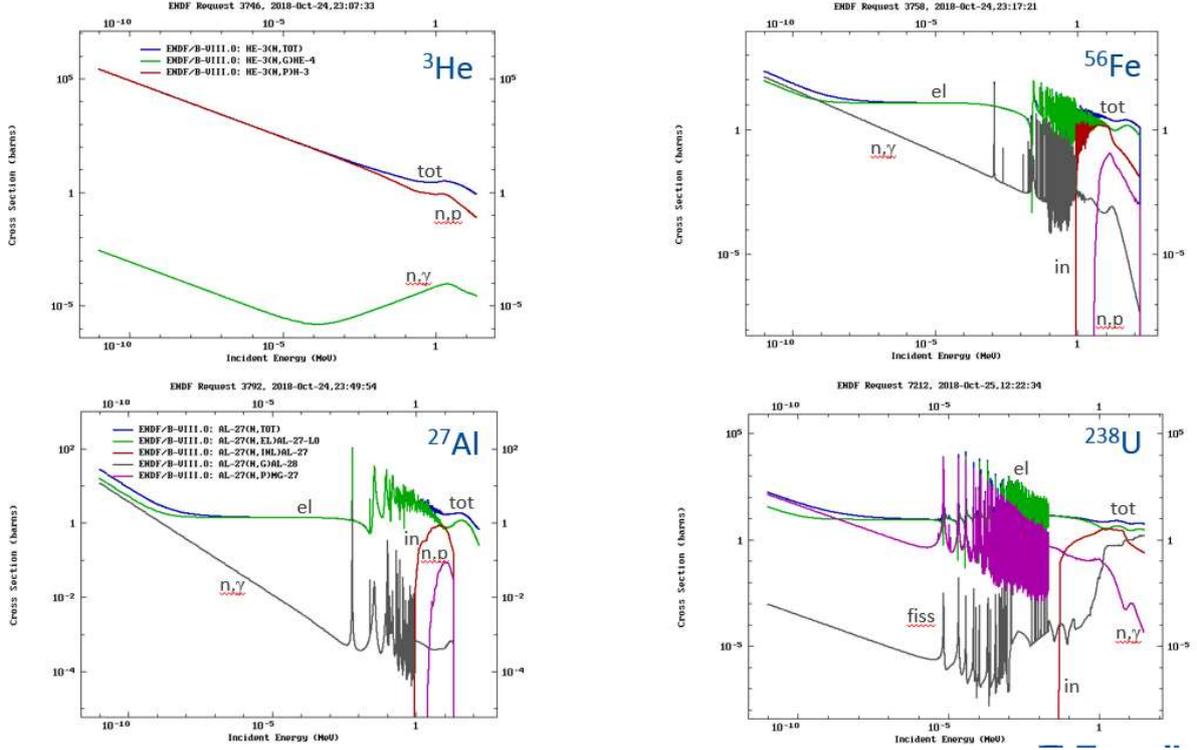

**Fig. 3:** Low-energy neutron cross-sections on $^3$He, $^{27}$Al, $^{56}$Fe and $^{238}$U, total and for different channels [17].

The second most important quantity to solve Eq. (1) or simulate particle transport by the Monte Carlo method is a double-differential cross-section:

$$\Sigma(\vec{r}, \Omega' \to \Omega, E' \to E) = \frac{d^2\sigma(i,j,A,E',E,\Omega',\Omega)}{dEd\Omega} \times \frac{N_A 10^{-27} \rho}{A}, \qquad (4)$$

where microscopic differential cross-section $\frac{d^2\sigma(i,j,A,E',E,\Omega',\Omega)}{dEd\Omega}$ is taken from pre-calculated databases, (e.g., ENDF [17] for low-energy neutrons), or some theoretical forms allowing - in simplest cases - even analytical solutions. At high-energies, a standard approach nowadays is to use *event generators*, performing Monte Carlo simulation through all the stages of particle interactions inside a nucleus like a quark-gluon cascade, hadron intranuclear cascade, pre-equilibrium stage, evaporation/fragmentation, and gamma-deexcitation. This is realized in DPMJET-III (Dual Parton Model) [18], FRITIOF (string model) [19], LAQGSM (quark-gluon string model) [20], RQMD (Relativistic Quantum Molecular Dynamics model) [21], and other theoretical models and codes. Microscopic differential cross-sections for pion and kaon production by protons on heavy nuclei are shown in Fig. 4. The LAQGSM model used in MARS15 [22, 23] reproduces experimental data very well. Before going further, the next subsection considers the important in accelerator applications concept of statistical weights and inclusive sampling.

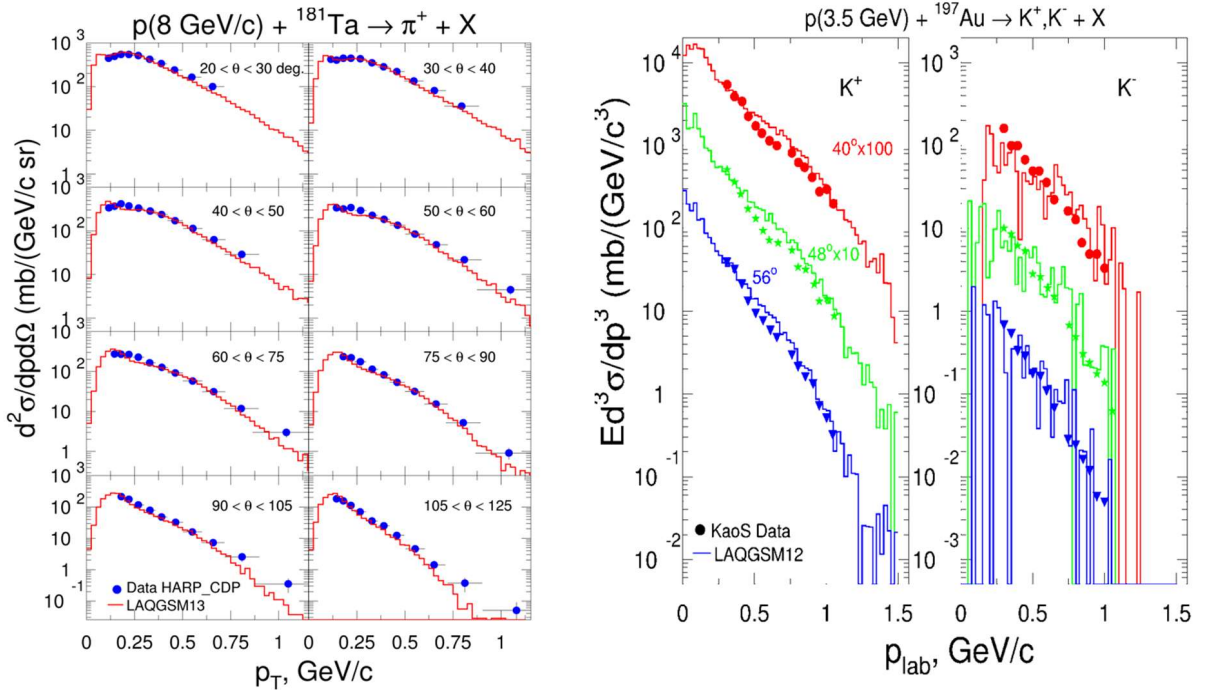

**Fig. 4:** Charged pion (left) and kaon (right) production differential cross-sections in interaction of 8 GeV/c and 3.5 GeV protons with tantalum and gold nuclei as calculated with LAQGSM *vs* experimental data [23].

*2.2.4 Statistical weights and inclusive sampling*

In contrast with the transport of low and intermediate energy hadrons through matter, a high-energy hadron cascade is a strongly branching process because of the multiple production of particles in the nuclear $hA$ interactions at energies above 10-30 GeV. Besides, the earlier theoretical models for high-energy particle-nuclear interactions suffered from a few inconsistencies. These factors – along with a limited computing power in 70's – forced developments of inclusive schemes for particle production in $hA$ interactions based on the method of *statistical weights* [3, 12, 13].

In its original simplest form, one can generate at each vertex only one "representative" sampled particle which carries a statistical weight W equal to the total secondary particle multiplicity at such a vertex. The functionals are calculated correctly on average if one sums such weights, with energy conserved in such an approach on average, but with correlations lost. Figure 5 illustrates this simplest scheme for the case of EMS analogous (exclusive) and weighted (inclusive) modelling. One can multiply and divide the integrand $\Sigma(x)$ in Eq. (1) by a function $p(x)$ of Eq. (2) to simplify the sampling and populate a region of interest:

$$\Phi = \int \Sigma(x)dx \implies \Phi = \int W(x)p(x)dx,$$

where $x = (E, \Omega)$ and $W(x) = \Sigma(x)/p(x)$ is a statistical weight. Any flux density functional in simulations is then calculated as

$$\Phi_N = \frac{1}{N}\sum_{k=1}^{N} W(x_k)\Delta l_k/\Delta V_k \qquad (5)$$

Here, $\Delta l_k$ is a segment of a particle trajectory in a detector volume $\Delta V_k$ in a $k$-history, and $N$ is a total number of histories. This approach is the basis of a version of the variance reduction techniques widely used in accelerator applications.

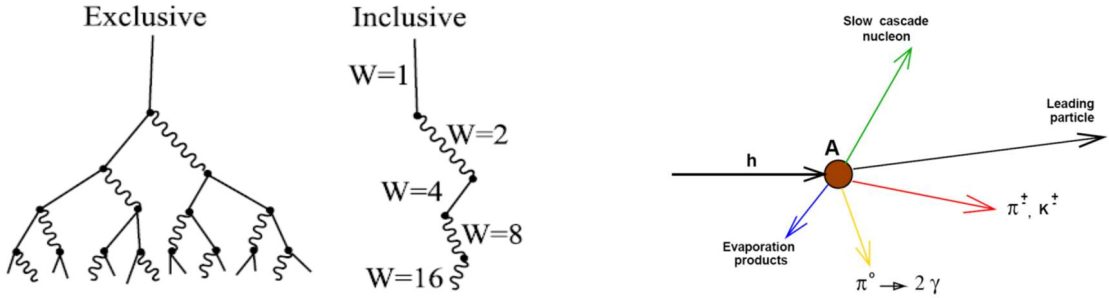

**Fig. 5:** The exclusive (left) and simplest inclusive (middle) schemes for modelling of electromagnetic showers (EMS). Efficient semi-inclusive scheme (right) for hA vertex modelling of early MARS versions.

At high energies, the hadron-nucleus (hA) vertex in simulations can be built to better match the needs of a specific application. Shown in Fig. 5 (right) scheme – developed in early versions of the MARS code [22] - provides a high efficiency $\varepsilon = (t\,\sigma^2)^{-1}$ in accelerator and shielding applications. Here $t$ is the CPU time needed to reach a statistical RMS error of σ. The statistical weights attached to each particle guarantee that the results are unbiased. A combination of exclusive, inclusive, semi-inclusive and hybrid sampling in the same session used in the current version of the MARS15 code can provide the highest efficiency $\varepsilon$ in many applications. Figure 6 shows particle tracks of EMS induced by one 10 GeV positron in a 3-cm tungsten slab followed by a 17-cm concrete slab. The showers were calculated with MARS15 in the exclusive mode (left) and a hybrid-20 mode (right). The latter means that the shower was simulated in the exclusive mode until the interaction vertices of a 20$^{th}$ level followed by the inclusive mode after that.

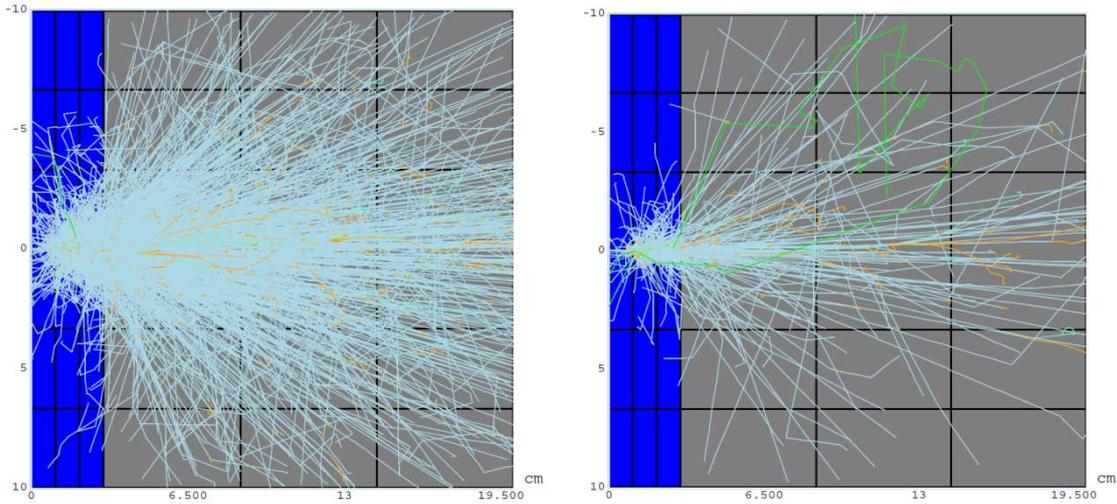

**Fig. 6:** Tracks of the 10-GeV EMS generated in the exclusive mode (left) and Hybrid-20 (mode). Track color ID: grey - photons, green – electrons and yellow – positrons.

*2.2.5   Simulation flow, response of additive detector and statistical accuracy*

Simulation of neutral stable particle interactions in a uniform block is straightforward. The range of a particle, $R$, before its discrete interaction is found by solving Eq. (1) whose right side is zero. In a coordinate system associated with the particle, its solution for neutral particles is $\Phi = \Phi_0\, exp(-\Sigma R)$. The corresponding probability density is $p(r) = \exp(-r)$, where $r = \Sigma R$. Then, the algorithm for simulating the range $R$ of neutral particles before their interaction: $r = -\ln(1-\gamma)$     or
$R = -\Sigma^{-1}\ln(1-\gamma)$. The radius-vector of the new interaction point can be determined now from the previous coordinates $\boldsymbol{r_0}$ as $\boldsymbol{r} = \boldsymbol{r_0} + R\boldsymbol{\Omega}$. The ratio of a specific channel cross-section to the total cross-section, gives us a probability of this channel to take place. By sampling it and applying a corresponding algorithm for sampling type, energy, and direction of generated secondary particles (database or event generator), we are equipped for simulation of processes initiated by a neutral particle in a uniform system.

Similar to the differential flux density and its functional definitions at the beginning of subsection 2.1, the *reading (response)* of any additive detector (in the broad sense) can be represented as

$$Res(t) = \iiint D(\vec{r},\vec{\Omega},E,t)\, \Phi(\vec{r},\vec{\Omega},E,t)\, d\vec{r}\, d\vec{\Omega}\, dE \tag{6}$$

where $D(\vec{r},\vec{\Omega},E,t)$ is the sensitivity function of the detector, (e.g., a light yield in scintillator). This is the average contribution to the detector readings from a unit path length of the particle with the coordinates $(\vec{r},\vec{\Omega},E,t)$ in the detector volume. Eq. (6) gives us the *alternative definition of the differential flux density* (with $D(x) = 1$) as a quotient of the sum of the particle track length segments $\Delta l_k$ in the spatial volume $\Delta V$ of a phase volume near a phase point $x = (\vec{r},\vec{\Omega},E,t)$, which is $\Phi(x) = \sum_k \Delta l_k /\Delta V$. Related *energy deposition* (with $D(x) = \Delta E$) for charged particles in $\Delta V$ is EDEP $(x) = \sum_k \Delta E_k \Delta l_k /\Delta V$. Both definitions are ready for prompt use in Monte Carlo *track length estimate* [3].

The particles produced at the discrete interaction vertex are placed into the history bank, with one of them taken for the further transport to a new interaction point using the techniques described. It is done with allowance for the particular features of the system (complex geometry, nonuniform material distribution, and composite materials), decay of unstable particles, possible quasi-continuous effects of the electromagnetic processes (ionization and radiative energy loss and multiple Coulomb scattering), and impact of magnetic and electric fields. Virtually any functional of the random quantities $\xi$ can be found directly during the simulation. The simulation of the history ends when the bank is empty, and all the particles are absorbed or emitted from the system. The simulation is then repeated $N$ times until the required statistical accuracy of the functionals is reached.

According to the central limit theorem, for large values of $N$, the distribution of the sum $\sum_{n=1}^{N} \xi_n$ is approximately normal. Therefore, the following relation is used to estimate the functional $\Phi$:

$$P\left\{abs\left(\frac{1}{N}\sum_{n=1}^{N} \xi_n - \Phi\right) < \delta\right\} \approx 0.997$$

This relation shows that – with a probability $P \cong 0.997$ - the error of the estimate is no greater than $\delta = 3\sqrt{D\xi/N}$, where $D\xi = M(\xi^2) - (M\xi)^2$, which is called a dispersion. That is, to calculate a statistical error of $M\xi$, one sums $\xi^2$ – in a course of Monte Carlo session – along with summing random quantities $\xi$.

### 2.2.6  Coulomb scattering, ionization energy loss and electromagnetic fields

To account for continuous (magnetic and electric fields) and quasi-continuous (multiple Coulomb scattering and ionization energy loss) processes in MCM, the charged particle path-length $x$ between its starting point and the next discrete interaction point or a boundary to the nearest adjacent physical region or a point of leakage from the absorber is subdivided into *steps s*, such that at every step the following typical conditions are fulfilled (*in the simplest approach*):

1. Angle due to multiple Coulomb scattering (MCS) is small ($\theta_{plane} \leq \sim 0.1$ mrad), (see Fig. 7).

2. Mean ionization energy loss in the continuous slowing-down approximation (CSDA) $\Delta E = abs\left(\frac{dE}{dx}\right) \times s \times \rho$ is small ($\frac{\Delta E}{E} \sim 1 - 5\%$). Here, $(-\frac{dE}{dx})$ is mean stopping power also called mass stopping power (see Fig. 8) and ρ is material density in $g/cm^3$.

3. Angle due to the effect of electro-magnetic field is small such that the field on the step is unchanged (from a practical standpoint) and the angle acquired by the particle on the step is miniscule (again, from the application-dependent point of view).

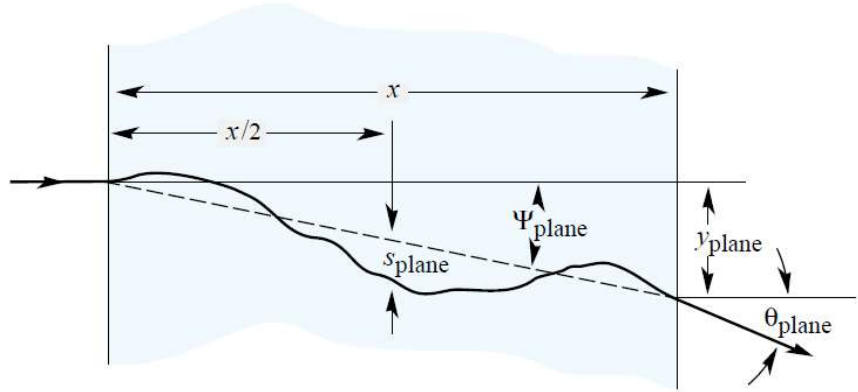

**Fig. 7:** Definition of parameters describing a trajectory of a charged particle in a slab [8].

In modelling multiple Coulomb scattering, it is sufficient for many applications to use a Gaussian approximation for the central 98% of the projected angular distribution with an RMS given by:

$$\theta_0 = \frac{0.0136 GeV}{\beta p} z \sqrt{\frac{x}{X_0}} [1 + 0.038 \ln(x z^2/X_0 \beta^2)] \tag{7}$$

where $z$, $\beta$ and $p$ are the particle charge number, velocity and momentum and $X_0$ is the medium radiation length. The value $\theta_0$ defined by Eq. (7) is related to the plane and spatial angles as

$$\theta_0 = \theta_{plane}^{rms} = \frac{1}{\sqrt{2}} \theta_{space}^{rms}$$

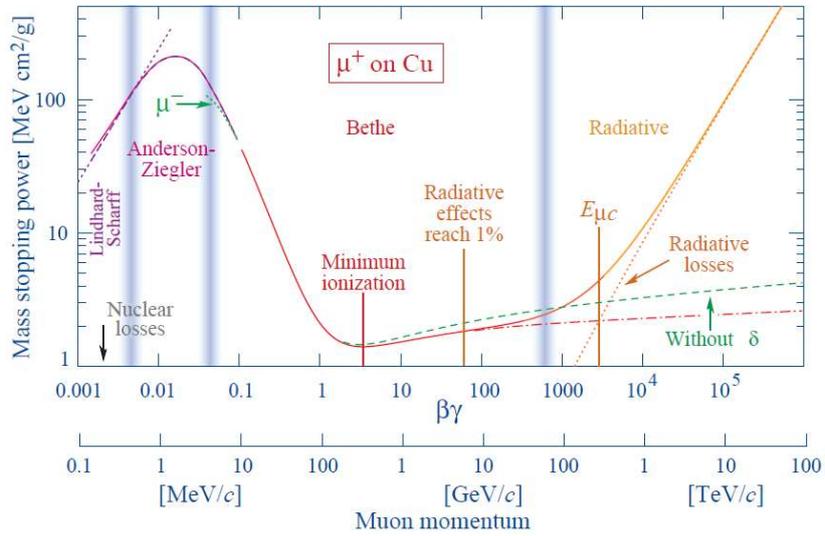

**Fig. 8:** Mass stopping power for positive muons in copper as a function of $\beta\gamma = p/Mc$ and muon momentum [8], with usual relations: $\gamma = \frac{1}{\sqrt{1-\beta^2}} = E_t/m$, $\beta\gamma = \sqrt{\gamma^2 - 1}$, $\beta = \frac{v}{c} = \frac{\beta\gamma}{\gamma} = p/E_t$

Mean stopping power for charged particles is described as $-\frac{dE}{dx} = a(E) + b(E) \times E$, where $a(E)$ is the electronic stopping power, $b(E)$ is due to radiative processes – bremsstrahlung, pair production and photonuclear interactions, and $E$ is particle energy [5]. Both $a(E)$ and $b(E)$ are slowly varying functions of energy at high energies. The electronic stopping power can be described as:

$$-\frac{1}{\rho}\frac{dE}{dx} = 4\pi N_A r_e^2 \, m_e c^2 z^2 \frac{Z}{A} 1/\beta^2 L(\beta) \qquad (8)$$

where $z$ is a projectile charge, $L(\beta) = L_0(\beta) + \sum_i \Delta L_i$, $L_0(\beta) = \ln(2\,m_e c^2 \beta^2 \gamma^2/I) - \beta^2 - \delta/2$ is the Bethe-Bloch formula, and $\sum_i \Delta L_i$. Several corrections $\Delta L_i$ are applied if a projectile is a not a single-charge particle, especially at low energies [24]: (i) Lindhard-Sørensen correction (exact solution to the Dirac equation; terms higher than $z^2$); (ii) Barkas correction (target polarization effects due to low-energy distant collisions); and (iii) shell correction. Projectile effective charge $z_{eff}$ comes separately as a multiplicative factor that takes into account electron capture at low projectile energies. For example, $z_{eff}$ is about 20 for 1-MeV/A $^{238}$U in Al, instead of a bare charge of 92. Moreover, the "cores-and-bonds" (CAB) method in MARS15 takes into account chemical bonds fitted to experiment for various compounds. Energy dependence of the mean stopping power (8) with corrections $\Delta L_i$ applied is shown in Fig. 9 for hydrogen, $^4$He, $^{12}$C, $^{27}$Al, $^{56}$Fe and $^{238}$U nuclei on silicon, in comparison with experimental data whenever available.

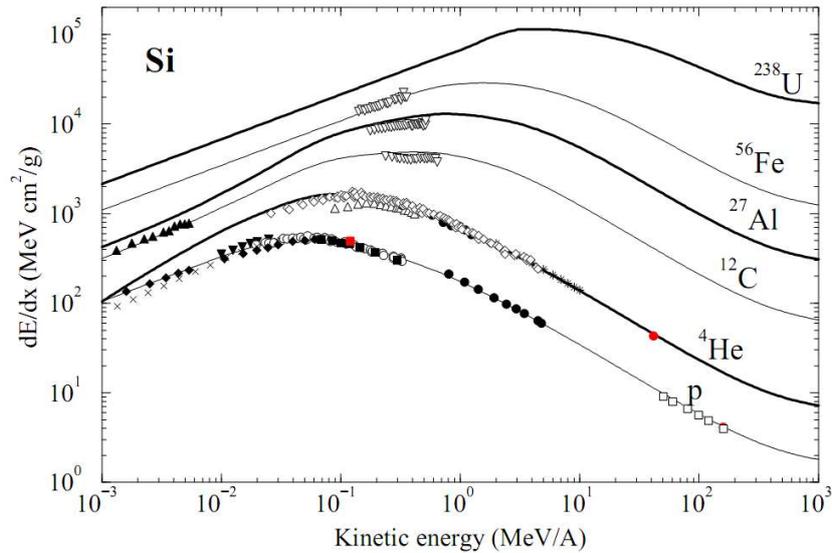

**Fig. 9:** Mean stopping power of projectiles from a proton to a bare uranium nucleus in silicon. Lines are calculations with Eq. 8 and symbols are experimental data referenced in [24].

The CSDA $dE/dx$ is widely used in quick estimations of energy loss by particle beams and in simplified simulations of energy loss and energy deposition along the charged particle tracks in hadronic and electromagnetic cascades. In a more sophisticated approach – used these days in several codes – precise modelling of knock-on electron production with energy-angle correlations taken into account is done for electronic losses. Radiative processes – bremsstrahlung, pair production, and inelastic nuclear interactions (via virtual photon) – for muons and high-energy hadrons are modelled, (e.g., in MARS15), using point-wise cross-sections [5]. Large ionization energy losses in "hard" collisions are also modelled as discrete interactions. The large-angle single Coulomb scattering is modelled via sophisticated algorithms. The latter is especially important for simulations in tiny accelerator and detector components. Customized steppers (*e.g.*, 8th order Runge-Kutta solver in MARS15) are used for charged particle tracking in complex geometry with complex magnetic and electric fields. Decays of unstable charged and neutral particles are modelled on a step either analogously or with one of the variance reduction techniques (modified decay length or forced decays). The Lorentz-invariant matrix element and polarization is taken into account in three-body decay kinematics.

If magnetic field ***B*** is present in a region where particle tracking and interactions are modelled, any charged particle will further change its direction upon passing the step *s*. If electric field ***E*** is present, the particle can in addition change its energy. These effects are governed by the Lorentz force ***F*** = $q[\boldsymbol{E} + \boldsymbol{v} \times \boldsymbol{B}]$, where $q$ and ***v*** are particle electric charge and velocity vector. If the step *s* is already small enough – due to ionization loss and MCS constraints – one can neglect the variations of ***B*** and ***E*** on the step. Then, the new direction **Ω** of the particle after the step is accurately derived from the equation of helical motion in a constant field. Otherwise – especially in complex geometry and field configurations - it is found by solving a high-order Runge-Kutta equation. The particle energy gain or loss in an electric field is calculated from the field component co-linear with the particle direction of motion and – in a case of a RF cavity – taking into account the distribution of the full ***E*** phase.

### 2.2.7 *Impact on materials*

Depending on the material, the level of energy deposition density, and the time structure, one can face a variety of effects in materials under irradiation. The physics behind each effect, conditions for the effect to take place and numerous examples are described in detail in Ref. [6, 7]. The two categories of materials response are related to the component lifetime and performance:
1. Component damage (lifetime):
   – thermal shocks, melting and quasi-instantaneous damage;
   – hydrodynamic tunneling with an intense beam drilling a hole in material of accelerator components [25, 26];
   – organic insulation property deterioration due to absorbed dose build-up;
   – radiation damage to inorganic materials due to atomic displacements amplified by helium and hydrogen production (see next subsection);
   – detector component radiation aging and damage.
2. Operational (performance):
   – superconducting magnet quenching;
   – soft errors (single-event effects specifically) in electronics;
   – detector performance deterioration;
   – radioactivation, prompt dose and impact on environment.

All these effects are accurately modelled with the codes described in Section 3, either directly or in concert with specialized codes such as ANSYS [27], MESA [28], SPHINX [29], LS-DYNA [30], and BIG2 [31].

### 2.2.8 *Displacements of atoms*

Atomic displacement cross-section $\sigma_d$ is a reference way to characterize the radiation damage induced by neutrons and charged particles in crystalline materials. To evaluate a number of displaced atoms, Norget, Torrens and Robinson proposed in 1975 a standard (so-called NRT-DPA) [32], which has been widely used since. DPA is the left side of Eq. (6) while $D = \Sigma_{DPA}$ is in the right side of Eq. (6). Energy of recoil fragments and new charged particles in - elastic and inelastic - nuclear interactions is used to calculate atomic displacement cross sections $\sigma_d$ for the NRT model – with or without Nordlund/Stoller damage efficiency ξ(T) – for a number of stable defects. Atomic screening parameters are calculated using the Hartree-Fock form-factors and recently suggested corrections to the Born approximation [33]. NJOY2016+ENDF/B-VIII.0(2018) system [17, 34] is used in MARS15 to generate the NRT/Nordlund/Stoller database for 490 nuclides for neutrons from $10^{-5}$ eV to 200 MeV; DPA in neutron-nuclear interactions above 200 MeV are treated the same way as for secondary charged particles and recoil fragments in nuclear interactions [6, 33, 35]. Atomic displacement cross-section $\sigma_d$ is calculated by integrating over recoil fragment energies $T_r$ and summing over all recoils generated in a given nuclear interaction:

$$\sigma_d = \sum_r \int_{E_d}^{T_r^{max}} \frac{d\sigma(E, Z_t, A_t, Z_r, A_r)}{dT_r} N_d(T_r, Z_t, A_t, Z_r, A_r) dT_r \tag{9}$$

where $N_d$ is the number of stable defects produced and $E_d$ is the displacement threshold. Similarly, a non-ionizing energy loss (NIEL) often-used in high-energy physics applications is calculated as:

$$\frac{dE}{dx}\bigg|_{ni} = N \sum_r \int_{E_d}^{T_r^{max}} \frac{d\sigma(E, Z_t, A_t, Z_r, A_r)}{dT_r} T_d(T_r, Z_t, A_t, Z_r, A_r) dT_r \tag{10}$$

where $N$ is a number of atoms per unit volume and $T_d$ is damage energy equal to the total energy lost in non-ionizing processes (atomic motion). The number of stable defects $N_d$ (or Frenkel pairs) is usually calculated from the NRT formula [33]

$$N_d^{NRT} = \frac{0.8}{2E_d} T_d \tag{11}$$

where $T_d$ is damage energy derived from the recoil fragment energy for given charge and atomic mass of irradiated material. Atomic displacement cross-section $\sigma_d$ (9) with the number of stable defects (11) is used in the course of a Monte Carlo session to calculate the value of DPA that is successfully applied to correlate data from many studies involving direct comparison from different irradiation environments. Nowadays, various corrections to the NRT model in the form $\xi(T) = \frac{N_d}{N_d^{NRT}}$ – called efficiency function – are used to account for atom recombination in elastic cascading. The most popular ones are the parametrization to the molecular dynamic calculations by Stoller [36] and the ARC-DPA (a thermal recombination-corrected DPA) proposed by Nordlund [37]. Atomic displacement cross-sections $\sigma_d$ for protons and neutrons on copper are shown in Fig. 10 as a function of projectile kinetic energy. One sees that the pure NRT model can overestimate the data by a factor of 2 to 3, while the use of Stoller and Nordlund physics-based corrections makes agreement very good.

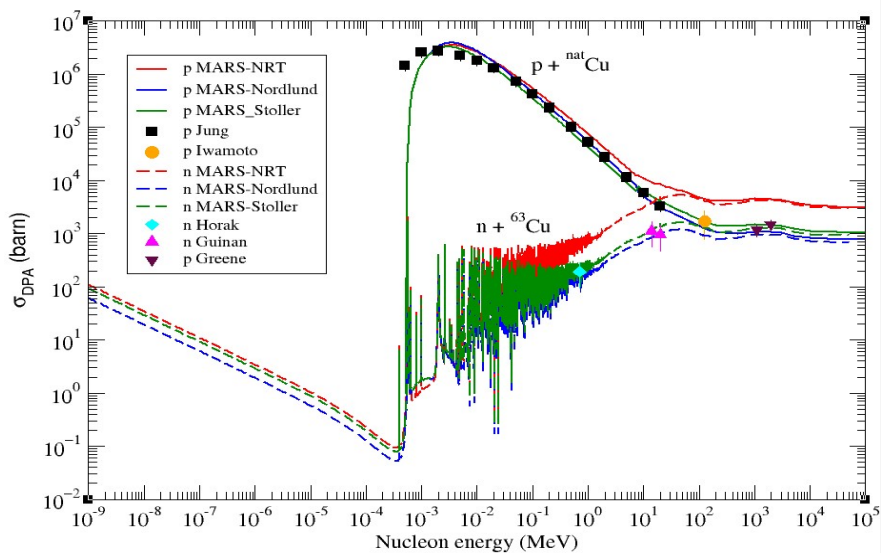

**Fig. 10:** Atomic displacement cross-section $\sigma_d$ for protons and neutrons on copper as calculated with three models in MARS15 in comparison with data [35].

At accelerators, radiation damage to inorganic structural materials—being primarily driven by displacement of atoms in a crystalline lattice —is amplified by increased hydrogen and helium gas production for high-energy beams. In the Spallation Neutron Source (SNS) beam windows, the ratio of He atoms to the number of displacements per target atom is about 500 times that in fission reactors. These gases can lead to grain boundary embrittlement and accelerated swelling. Responsible for hydrogen and helium gas generation, proton and α-particle production is promptly modelled in the simulation codes described in the next section.

## 3   Simulation of particle-matter interactions and transport in modern Monte Carlo codes

As stated in Ref. [38], nowadays the use of general-purpose particle interaction and transport Monte Carlo codes is the most accurate and efficient choice for assessing impact and consequences of particle-matter interactions at accelerators. Due to the vast spread of such codes to all areas of particle physics and the associated extensive benchmarking with experimental data, the modelling has reached an unprecedented accuracy. Furthermore, most of these codes allow the user to simulate all aspects of a high energy particle cascade in one and the same run: from the first interaction of a primary beam (of up to TeV energies) over the transport and re-interactions (hadronic and electromagnetic) of the produced secondaries, to detailed nuclear fragmentation, the calculation of radioactive decays, secondary electromagnetic showers, muon and neutrino generation and their interaction with surroundings.

Besides the physics processes and quantities described in the previous sections, simulation codes considered in this paper can provide the following quantities – directly calculated in the course of a Monte Carlo session – to derive the materials response: spatial distributions of energy deposition density and heat load to directly evaluate absorbed dose, proximity to the quench limit in superconducting magnets, instantaneous temperature rise and maps to feed ANSYS for thermal and stress analyses, particle fluence, star density (density of inelastic nuclear interactions responsible for nuclide production above 30 MeV), DPA, and prompt and residual effective dose. Also, they can provide numerous functionals integrated over pre-defined regions such as power dissipation, energy and angular spectra, total isotope production, time-of-flight distributions, and many other either built-in or user-specified quantities and distributions. Hydrodynamic and thermal-stress analysis (ANSYS) codes as well as radiological ES&H tools can use those quantities as input to evaluate the material response.

Multithreading is a common technique used by particle-matter interaction codes in CPU-hungry applications. It is applied to one process to enable parallel execution on a multiprocessing system. Multithreading is a widespread programming and execution model that allows multiple threads to exist within the context of one process. These threads share the process's resources but are able to execute independently. Multithreading is a user-friendly alternative to a multiple-core fail-proof approach (used, for example, in MARS for decades) with thousands of independent jobs on a cluster submitted with a user-created script and with results averaged at a post-processing stage.

Extending a brief overview of the simulation codes [38], an account of the current versions of five widely used codes FLUKA, GEANT4, MARS15, MCNP6, and PHITS with examples of their use in accelerator applications and results of recent benchmarking is given in this section.

### 3.1   FLUKA

FLUKA [39-41] is a general-purpose particle interaction and transport (Fortran-**77**) code. It comprises all features needed for radiation protection, such as detailed hadronic and nuclear interaction models up to 10 PeV, full coupling between hadronic and electromagnetic processes and numerous variance reduction options. The latter include weight windows, region importance biasing, and leading particle, interaction, and decay length biasing (among others). The capabilities of FLUKA are very good for studies of induced radioactivity, especially with regard to nuclide production, decay, and transport of residual radiation [38]. In particular, particle cascades by prompt and residual radiation are simulated in parallel based on the microscopic models for nuclide production and a solution of the Bateman equations for activity build-up and decay. FLUKA is the *de facto* the official code in numerous LHC and other applications at CERN.

The highest priority in the design and development of FLUKA has always been the implementation and improvement of sound and modern physical models. Microscopic models are adopted whenever possible, consistency among all the reaction steps and/or reaction types is ensured, conservation laws are enforced at each step, and results are checked against experimental data at the single interaction level. As a result, final predictions are obtained with a minimal set of free parameters fixed for all energy/target/projectile combinations. Therefore, results in complex cases, as well as properties and scaling laws, arise naturally from the underlying physical models, predictivity is provided where no experimental data are directly available, and correlations within interactions and among shower components are preserved. FLUKA can handle very complex geometries, using an improved version of the well-known Combinatorial Geometry (CG) package. The FLUKA CG has been designed to also correctly track charged particles, even in the presence of magnetic or electric fields. Various visualization and debugging tools are also available. Similar to the MARS15 code, FLUKA has a double capability to be used in a biased mode as well as a fully analogue code. That means that while it can be used to predict fluctuations, signal coincidences and other correlated events, a wide choice of statistical techniques are also available to investigate punch-through or other rare events in connection with attenuations by many orders of magnitude.

A detailed FLUKA model of the LHC betatron collimation region with all the geometry details, materials and magnetic fields taken into account is shown in Fig. 11. Comparison of simulation results in this region will be shown in the benchmarking sub-section.

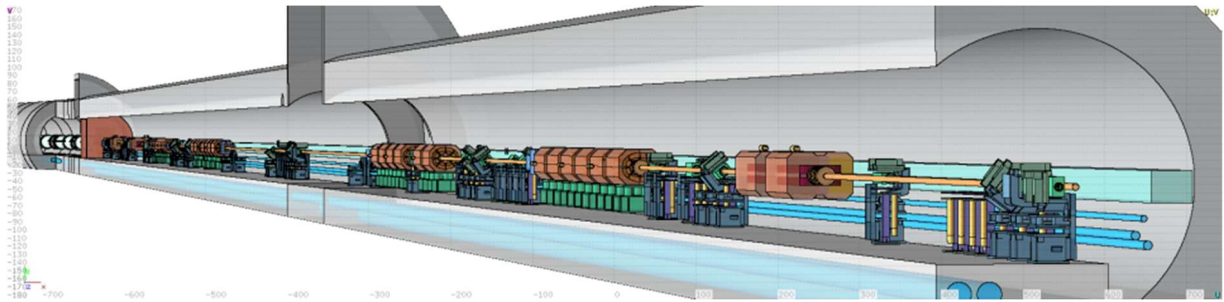

**Fig. 11:** FLUKA geometry model of the LHC betatron cleaning insertion.

### 3.2 GEANT4

GEANT4 [42-45] is an object-oriented toolkit consisting of a kernel that provides the framework for particle transport, including tracking, geometry description, material specifications, management of events and interfaces to external graphics systems. The kernel also provides interfaces to physics processes. It allows the user to freely select the physics models that best serve the particular application needs. Implementations of interaction models exist over an extended range of energies, from optical photons and thermal neutrons to high-energy interactions required for the simulation of accelerator and cosmic ray experiments. The code is the industry standard for HEP detector simulation. To facilitate the use of variance reduction techniques, general-purpose biasing methods such as importance biasing, weight windows, and a weight cut-off method have been introduced directly into the toolkit. Other variance reduction methods, such as leading particle biasing for hadronic processes, come with the respective physics packages [38].

A comprehensive set of the well-established models comprises GEANT's physics lists for users to choose from. Substantial efforts were and are made by the GEANT4 team on validation and verification of electro-magnetic physics in the code and hadronic physics loosely defined to cover any reaction which can produce hadrons in final state: purely hadronic interactions, lepton- and gamma-induced nuclear reactions, and radioactive decay. Models and cross-sections which span an energy range from sub-eV to TeV are provided. Following the toolkit philosophy, more than one model or process is usually offered in any given energy range in order to provide alternative approaches for different applications. GEANT4's performance was noticeably improved after several international benchmarking campaigns over last 15 years.

There are several ways to build a geometry model in GEANT4. The standard one is to write C++ code that contains all the definitions, materials, dependence, position and hierarchy assignments, and arranges all these in the model. The shapes or geometrical primitives can be taken from the toolkit's built-in comprehensive library (see Fig. 12). Fragments of the model can be imported and exported from external files according to two different formats, GDML or plain ASCII text. GEANT4 provides internal modules which allow the interpretation and conversion of these formats to and from the internal geometry representation, without the need for C++ programming for the implementation of the various detector description setups. Figure 13 shows examples of GEANT4 viewers.

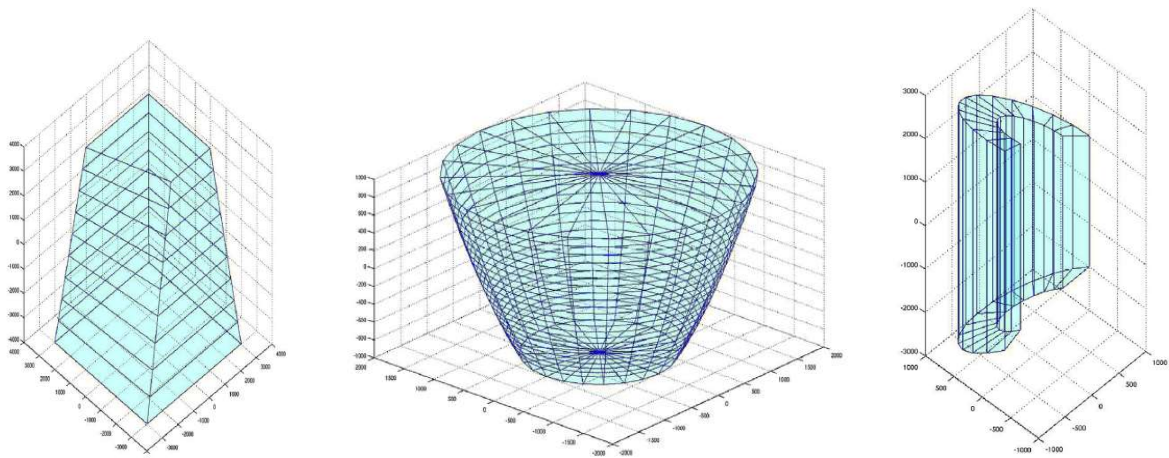

**Fig. 12:** Trapezoid (left), parabolic solid (center), and cut tube (right) from GEANT4 geometrical primitive library

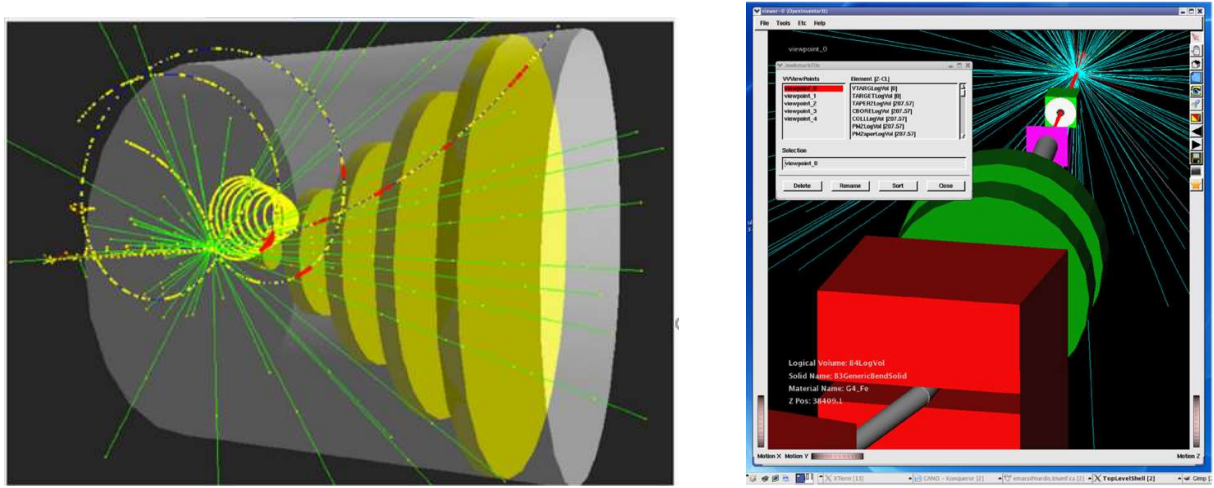

**Fig. 13:** Snapshots from GEANT4 OpenGL viewer wrapped in Qt (left) and Open Inventor extended viewer (right).

## 3.3 MARS15

MARS15 [16, 22, 23, 46-48] is a set of Fortran-77 and C++ programs for Monte Carlo simulations of coupled hadronic and electromagnetic cascades, with heavy ion, muon, and neutrino production and interaction. It covers a wide energy range: 1 keV to 100 TeV for muons, hadrons, heavy ions, and electromagnetic showers and $10^{-5}$ eV to 100 TeV for neutrons. Nuclear interactions as well as practically all other strong, weak and electromagnetic interactions in the entire energy range can be simulated either inclusively or exclusively i.e., in a biased mode or in a fully or partially analogue mode. Nuclide production, decay, transmutation, and calculation of the activity distribution is done with the built-in DeTra code [49]. MARS15 uses ENDF/B-VIII.0(2018) nuclear data to handle interactions of neutrons with energies below 14 MeV and derive the NRT/Stoller/Nordlund DPA x-sections below 200 MeV. The elemental distributions are automatically unpacked into isotope distributions for both user-defined and those from the 172 built-in materials. A tagging module allows one to tag the origin of a given signal for source term or sensitivity analyses. Several variance reduction techniques such as weight windows, particle splitting, and Russian roulette are possible.

Several ways are offered to describe geometry models to a MARS15 user: a basic solid body representation, a ROOT-based engine [50], GDML files, MARS input files generated with the G4 beamline Bruit De Fond [51], and Fortran or C++ codes written by a user according to templates provided. For simulation in accelerator environment, the MARS-MAD Beamline Builder (MMBLB) and active merge with PTC tracking within MAD-X [52] are used for a convenient creation of accelerator models and multi-turn tracking coupled to cascade simulation in accelerator and beamline lattices [53]. Fragments of the MARS15 geometry models are shown in Figs. 14 and 15 for the Higgs Factory muon collider, J-PARC 3-GeV ring, Fermilab Booster and LHC IP5 interaction region. MARS15 is routinely used in concert with ANSYS for iterative studies of thermo-mechanical problems and can be interfaced to a hydrodynamic code to study phase transition and "hydrodynamic tunneling" – first done at SSC for a 20-TeV proton beam in 1993 [25].

Developed over years, the MARS15 model of the Long Baseline Neutrino Facility (LBNF) [54] to feed the DUNE experiment [55] is used to optimize the LBNF design as well as verify the neutrino spectra at the DUNE Near and Far detectors. The model starts at the extraction from the Main Injector 300 meters upstream of the LBNF target and continues through a primary beamline to the target station followed by a decay channel with a hadron absorber complex at 220 meters from the target, with a neutrino Near Detector at about 450 meters from the target. The model is built of almost $10^5$ elements and 120 materials. In routine applications, the model is reduced to about 17,000 elements. As an example, Fig. 16 shows the target station model with particle tracks and MARS15-ANSYS calculated temperature profiles in the hottest aluminum block of the hadron absorber.

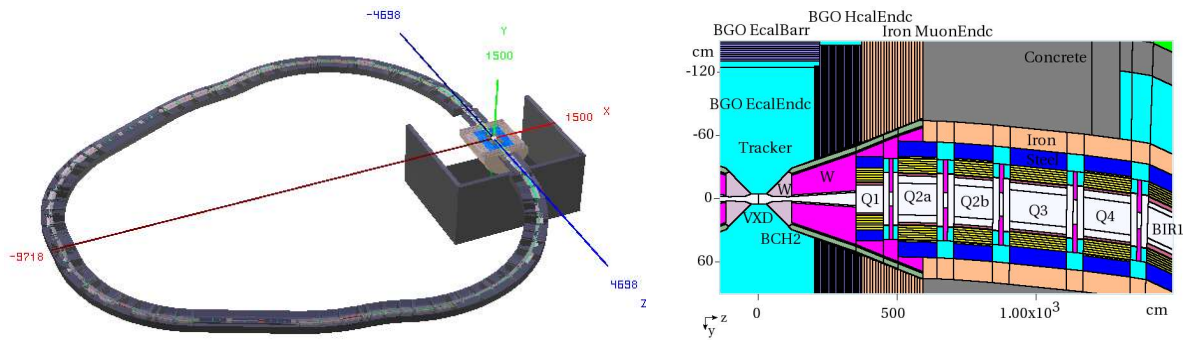

**Fig. 14:** MARS15 model of the Higgs Factory muon collider 300-m ring with SiD-like detector (left) and Machine-Detector Interface (right).

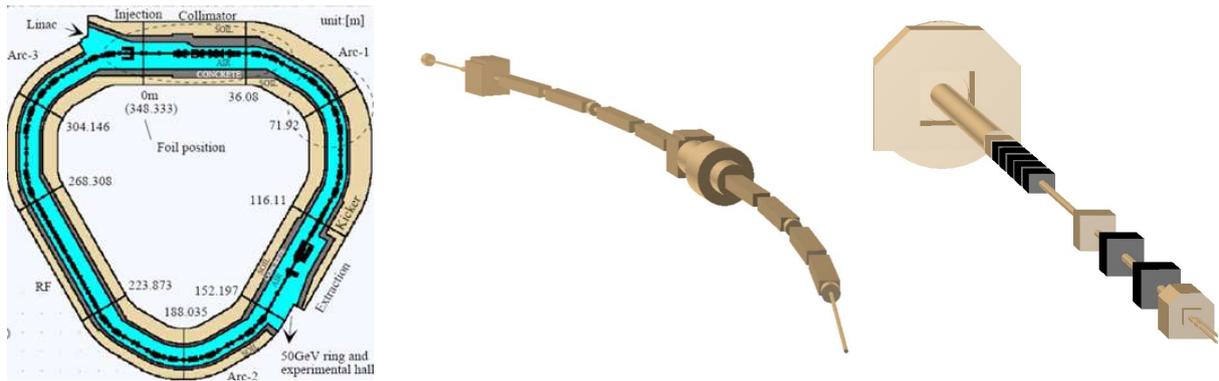

**Fig. 15:** MARS15 models of the J-PARC 3-GeV 350-m long in circumference proton accelerator (left), Fermilab Booster collimation region (center) and LHC IP5 region (right).

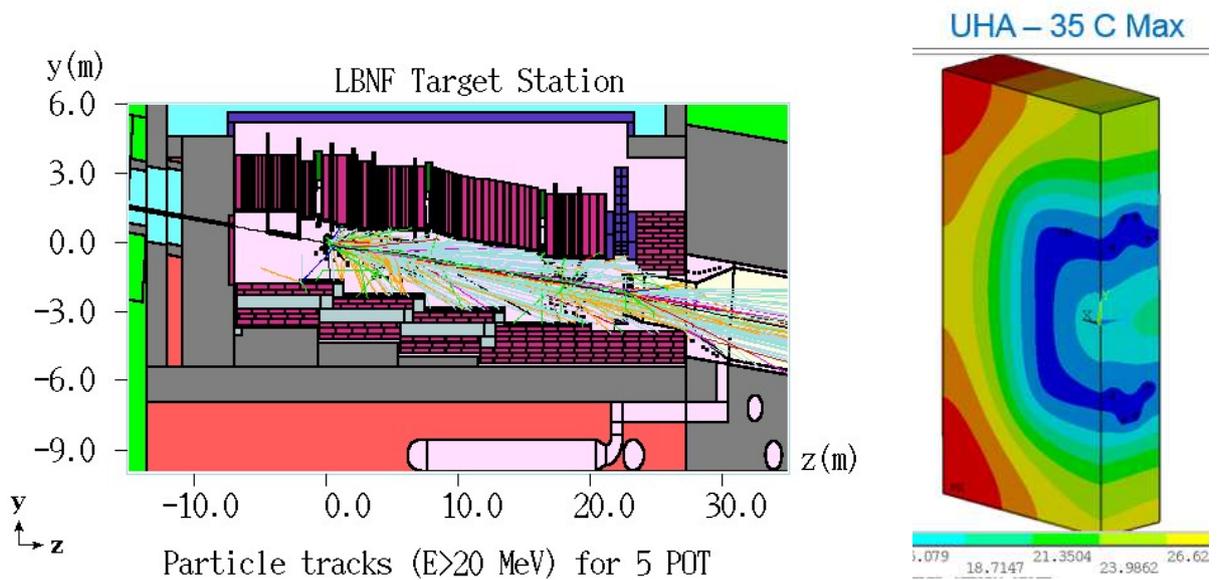

**Fig. 16:** MARS15 geometry model of the LBNF target station with particle tracks shown for 5 protons on target (left). Temperature profile calculated with MARS15-ANSYS in the hottest aluminum block of the LBNF Hadron Absorber at 220 m from the target (right).

### 3.4 MCNP6

MCNP6 [56-58] is the latest version of the Monte Carlo N-Particle transport (MCNP) family of particle interaction and transport codes (Fortran-90) and features comprehensive and detailed descriptions of the related physical processes. It transports 37 different particle types, including ions and electromagnetic particles. The particle interaction and transport modules use standard evaluated data libraries mixed with physics models where such libraries are not available. The code is considered by many as the industry standard for simulation in reactor, medical, space and low- and medium-energy accelerator applications. The transport is continuous in energy. MCNP6 contains one of the most powerful implementations of variance reduction techniques. Spherical mesh weight windows can be created by a generator in order to focus the simulation time on certain spatial regions of interest. In addition, a more generalized phase space biasing is also possible through energy- and time-dependent weight windows. Other biasing options include pulse-height tallies with variance reduction and criticality source convergence acceleration [38].

Geometry options in MCNP6 include traditional surface-based, voxel lattice, constructive solid and unstructured mesh. Figure 17 shows an FRIB pre-separator model built for MCNP/PHITS simulations [59]. Magnetic field implementation includes: (1) Constant dipole, square-edge quadrupole and quadrupole with a fringe-field kick, all in low-density materials, such as air, and (2) COSY maps only in vacuum and specific to one particle type. Both are rather limited compared to four other codes considered in this section with the arbitrary EM field capability in arbitrary geometry/materials. The unique feature is that MCNP6 is considered risk level two software (death is risk level one). Meaning, it is treated as if failure of the software could result in temporary injury or illness to workers or the public. Therefore, it provides a set of hundreds of automated verification, validation, and regression tests. The latter is for detecting unintended changes to the code and installation testing. The optional feature in MCNP6 is super-precise simulation of EMS in the energy range of 1 eV to 100 GeV. Contrary to the standard condensed-history modelling of electron transport, this mode uses a super-accurate single-event approach which is naturally extremely CPU-time consuming.

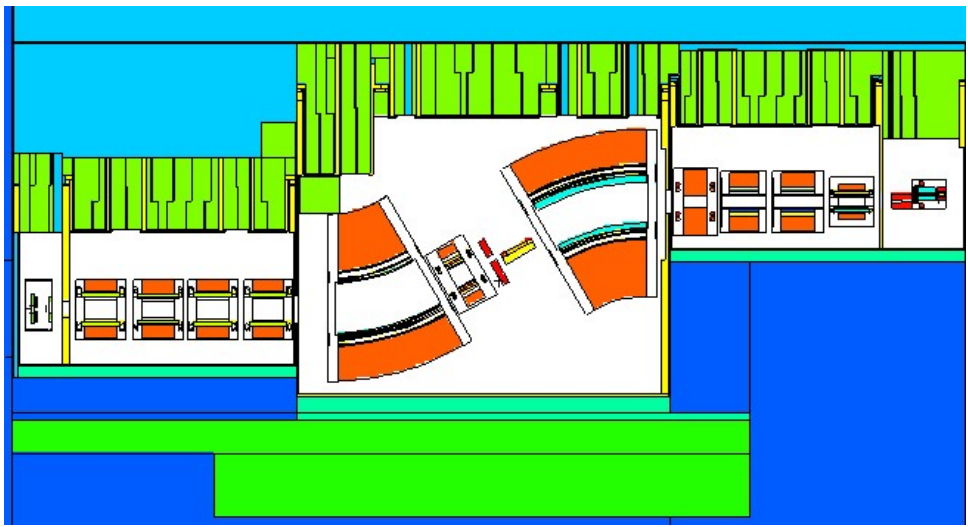

**Fig. 17:** Radiation transport model of the FRIB pre-separator vacuum vessels built with MCAM [60] and VISED [61] and used for Monte Carlo calculations [59] with MCNPX and PHITS.

### 3.5 PHITS

PHITS [62-64] is the Particle and Heavy-Ion Transport code System (Fortran-77). It was among the first general-purpose codes to simulate the transport and interactions of heavy ions in a wide energy range, from 10 MeV/nucleon to 100 GeV/nucleon. It is based on the high-energy hadron transport code NMTC/JAM that was extended to heavy ions. The transport of low-energy neutrons employs cross sections from evaluated nuclear data libraries such as ENDF and JENDL below 20 MeV. Electromagnetic interactions are simulated based on the EGS5 code in the energy range between 1 keV and 100 MeV for electrons and positrons and between 1 keV and 100 GeV for photons. Several variance reduction techniques, including weight windows and region importance biasing, are available [38]. An accurate calculation of DPA is supported by dedicated experiments with medium-energy protons.

The geometrical configuration of a simulation is set with general geometry (GG) in a manner similar to MCNP (see Figs. 17 and 18). The interactive solid modeler Simple-Geo (FLUKA) can be used for generating the geometries written in PHITS-readable GG format. Geometries based on Computer-aided Design (CAD) can be incorporated into PHITS by converting CAD data into tetrahedral-mesh geometries. In addition, CAD geometries can be directly converted into the PHITS-readable GG format by using SuperMC [65]. Electromagnetic fields and gravity can be considered in the transport simulation of all particles. The time evolution of radioactivity is estimated by a built-in DCHAIN-SP module.

### 3.6 Simulation code recent benchmarking

At the CERN High-Energy Accelerator Mixed Field facility CHARM, there was a recent code benchmarking campaign on thermal neutron fields induced by the 24 GeV/c proton beam on a 50-cm thick copper target [66]. Its schematic view is shown in Fig. 18. Experimental data was compared to the results of calculations by the PHITS, FLUKA, and MARS15 codes. As concluded in [66] and seen in Fig. 19, PHITS results agree with the data within 50%, while FLUKA, MARS15 and PHITS results agree with each other within 30%.

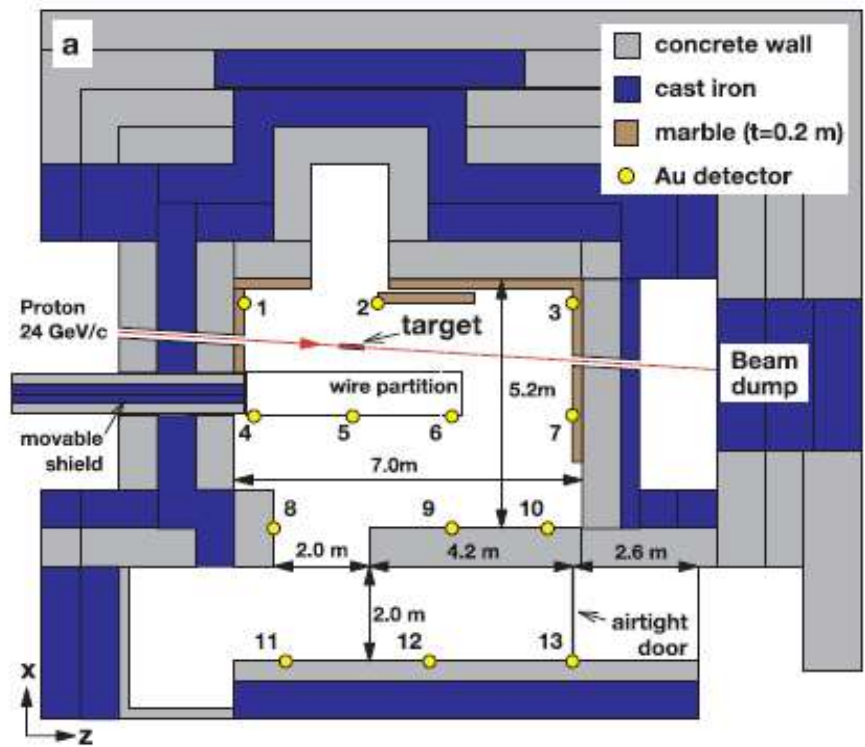

**Fig. 18:** A cross-sectional view taken along the Cu target plane of the CHARM facility [63]. The numbers 1–13 indicate the experimental locations of the gold foils.

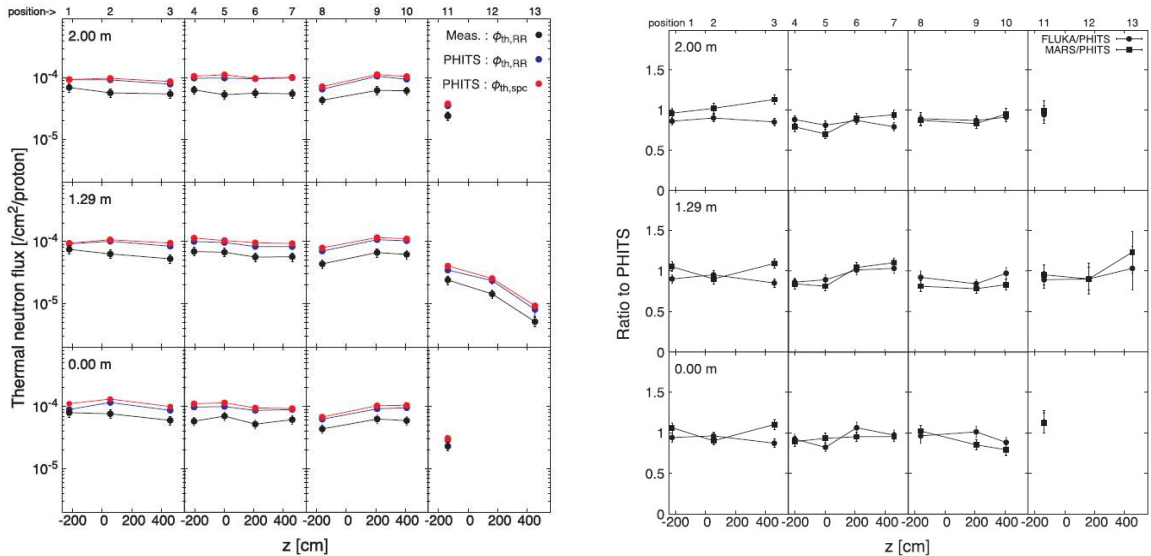

**Fig. 19:** Experimental (black symbols) and PHITS calculation (blue and red symbols) results for the thermal neutron flux (left). Ratios of the FLUKA (circles) and MARS (squares) results to PHITS results for the thermal neutron flux calculated at 3 heights *vs* longitudinal position in the CHARM facility (right) [63].

To get more confidence in the MARS15-based LBNF target station design, a benchmarking campaign on air activation has been recently undertaken at the Fermilab NuMI target station for a 120-GeV beam on target [67]. The results of comparison are shown in Table 1. MARS15 underestimates the $^{41}$Ar production rate by 50% (which is not that bad for this very difficult dynamic benchmarking) and agrees with data for other nuclides within 10-30%.

**Table 1:** Measured and calculated production rates (cm$^{-3}$ POT$^{-1}$ s$^{-1}$) for the most important radionuclides generated in the air in the beam enclosure of the NuMI target chase [67]

|  | $^{41}$Ar | $^{11}$C | $^{13}$N | $^{15}$O |
|---|---|---|---|---|
| Exp. data | 1.98×10$^{-12}$ | 6.38×10$^{-11}$ | 4.07×10$^{-11}$ | 3.50×10$^{-11}$ |
| Fermilab ES&H methodology | 6.85×10$^{-12}$ | 2.22×10$^{-10}$ | 5.22×10$^{-11}$ | 9.16×10$^{-11}$ |
| MARS15 | 1.08×10$^{-12}$ | 4.44×10$^{-11}$ | 3.71×10$^{-11}$ | 4.16×10$^{-11}$ |
| MARS15/data | 0.55 | 0.70 | 0.91 | 1.19 |

As described in Ref. [7], several quench tests were performed at LHC to explore the actual quench limits of the superconducting magnets and verify the quality of theoretical calculations [68]. In particular, during a so-called collimation quench test at a beam energy of 4 TeV, the horizontal primary collimator of the betatron cleaning insertion (see Fig. 11) was impacted by a peak proton loss rate equivalent to about 1 MW for 1 s, with no quench occurring in the downstream dispersion suppressor. The propagation of the induced particle shower was measured by the beam loss monitor system, giving a picture of the energy deposition profile over several hundred metres, as shown in Fig. 20. This provided a very challenging opportunity to benchmark the adopted SixTrack-FLUKA simulation chain [69], which yielded the impressive agreement reported in the figure, both in terms of pattern and absolute signal comparison, spanning a few orders of magnitude.

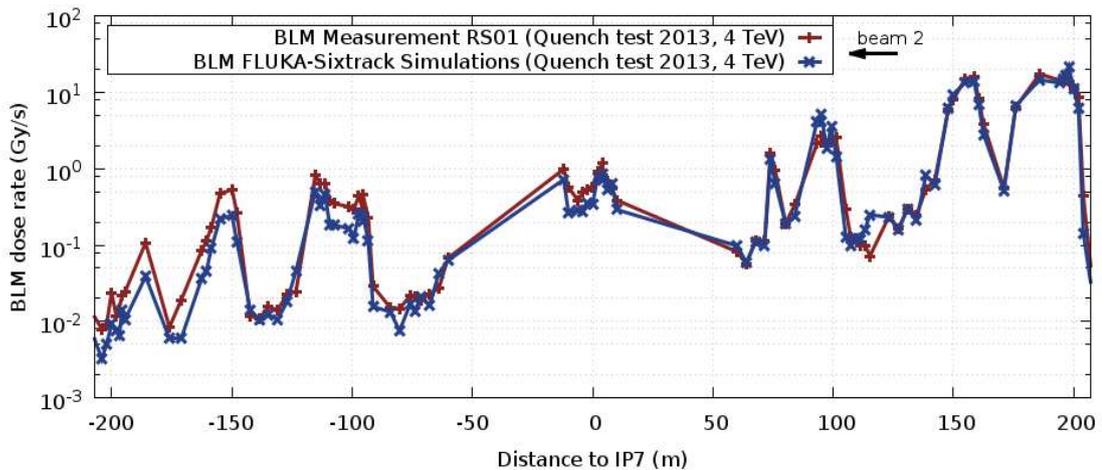

**Fig. 20:** Absolute beam loss monitor signal pattern at the peak loss rate of the 2013 LHC collimation quench test, averaged over the shortest available time interval of 40 μs (RS01): data (red) are compared with predictions (blue) by the SixTrack-FLUKA coupling according to the simulation strategy discussed in Ref. [66].

## 3.7 Active merge of particle-matter interaction and accelerator tracking codes

In accelerator applications, one provides – one way or another - a cross-talk between an accelerator tracking code and a code for particle-matter interactions. For example, the MARS code was linked for more than two decades with a STRUCT code [70]. There has been a quantum leap in coupling general-purpose code with tracking code for accelerators:

- MMBLB = MAD-MARS Beam Line Builder (since 2000) [71]. For more than two decades, the accelerator tracking was done with the STRUCT code [70]. Recently, MARS15 was switched to the ROOT-based beam-line builder and integrated with the MAD-X system [52].
- BDSIM combines C++ in-vacuum accelerator style particle tracking and GEANT4 physics (since the mid-2000s) [72].
- FLUKA LineBuilder and Element Database [73] and active coupling to SixTrack [74]; the two codes communicate with each other through a network port [75].

In the MARS15-MAD-X system with PTC modules, a library containing functions and C++ classes which interfaces MARS and MAD-X is now packed with the MARS15 distribution. The library allows to:

- Create a 3-D TGeo ROOT geometry model for the sequence described in a MAD-X input file. Alignment of elements is performed by means of the MAD-X survey table.
- Define transformation for each point in the phase space used in MAD-X-PTC to the phase space used in MARS15 and vice versa.
- Inject particles transported by MARS15 to MAD-X-PTC using a formulated acceptance for the accelerator code model.
- For particles transported in MAD-X-PTC, perform a check of boundary crossing against the ROOT geometry in MARS15; the particle is forwarded to the MARS15 stack.

The original MAD-X code available at Fermilab was modified to allow a particle to start from the upstream end of an arbitrary element in the sequence and check the aperture crossing against the MARS15 geometry not only at the entrance and exit of the element, but also all along a curved track (e.g., in dipoles). Fig. 21 shows the results of a successful application of this system in the design of the beam collimation system for Fermilab 8-GeV Recycler.

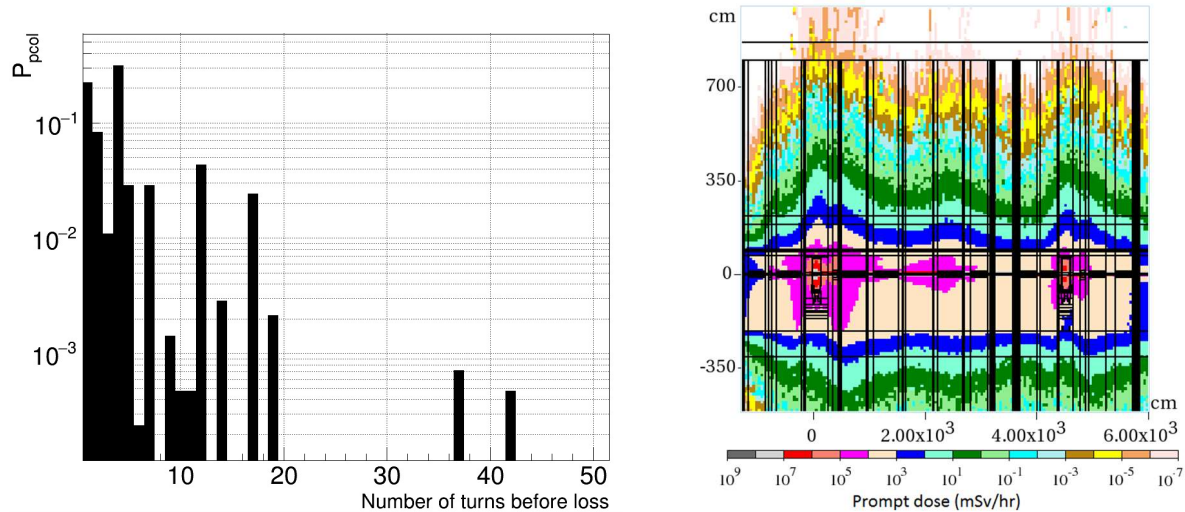

**Fig. 21:** Probability for beam halo protons passed through the primary collimator be intercepted in the Fermilab Recycler collimation system as a function of turns (left). Prompt dose distribution in the Recycler collimation region (right).

**Acknowledgements**

I wish to thank I. Rakhno, S. Striganov, and I. Tropin for their contributions. This document was prepared using the resources of the Fermi National Accelerator Laboratory (Fermilab), a U.S. Department of Energy, Office of Science, HEP User Facility. Fermilab is managed by Fermi Research Alliance, LLC (FRA), under Contract No. DE-AC02-07CH11359.

**References**

[1] B. Rossi, *High Energy Particles*, (Prentice-Hall, Inc., Englewood Cliffs, NJ, 1952).

[2] V.S. Barashenkov and V.S. Toneev, *Interaction of High-Energy Particles and Nuclei with Nuclei* (Atomizdat, Moscow, 1972) [in Russian].

[3] A.N. Kalinovskii, N.V. Mokhov, and Yu.P. Nikitin, *Passage of High-Energy Particles through Matter* (American Institute of Physics, New York, 1989).

[4] A. Ferrari and P.R. Sala, The Physics of High Energy Reactions, *Proc. Workshop on Nuclear Reaction Data and Nuclear Reactors Physics, Design and Safety*, Trieste, April 1996, Ed. A. Gandini and G. Reffo (1998), p. 424.

[5] D.E. Groom, N.V. Mokhov, and S.I. Striganov, *Muon Stopping-Power and Range Tables: 10 MeV – 100 TeV* (Atomic Data and Nuclear Data Tables **78**, 2001) 183-356.

[6] N.V. Mokhov, *Rev. Accel. Sci. Technol.* **6** (2013) 275-290. http://dx.doi.org/10.1142/S1793626813300132.

[7] N.V. Mokhov and F. Cerutti, *Proc.Joint Int. Accelerator School*, Newport Beach, US, Nov. 2014, ed. R. Schmidt, CERN-2016-002 (2016) 83-110.

[8] H. Bishel, D.E. Groom, and S.R. Klein, *Review of Particle Physics*, Phys. Rev. **D98**, 030001 (2018) 446-460.

[9] R.G. Alsmiller and F.S. Alsmiller, A Perturbation Method for Solving the Angle-dependent Nucleon-meson Cascade Equations, ORNL Report No. ORNL-3467 (1963).

[10] F.S. Alsmiller, A General Category of Soluble Nucleon-meson Cascade Equations, ORNL Report No. ORNL-3746 (1965).

[11] K. O'Brien. Nucl. Instrum. Methods **72** (1969) 93.

[12] J. Spanier and E.M. Gelbard, *Monte Carlo Principles and Neutron Transport Problems* (Addison-Wesley, Reading, 1969).

[13] I.M. Sobol, *Numerical Monte Carlo Methods* (Nauka, Moscow, 1973) [in Russian].

[14] A.M. Kolchuzhkin and V.V. Uchaikin, *Introduction to the Theory of the Passage of Particles Through Matter* (Atomizdat, Moscow 1978) [in Russian].

[15] *Review of Particle Physics*, Phys. Rev. **D98**, 030001 (2018) 600.

[16] N.V. Mokhov and S.I. Striganov, *AIP Conf. Proc.* **896** (2007) 50. http://dx.doi.org/10.1063/1.2720456.

[17] *ENDF/B-VIII.0 Evaluated Nuclear Data Library*, https://www.nndc.bnl.gov/endf/b8.0/.

[18] S. Roesler, R. Engel, and J. Ranft, The Monte Carlo Event Generator DPMJET-III, *Advanced Monte Carlo for Radiation Physics, Particle Transport Simulation and Applications*. ISBN 978-3-642-62113-0. Springer-Verlag Berlin Hiedelberg, 2001, p. 1033.

[19] B. Anderson, G. Gustafson, and Hong Pi, Zeitschrift fur Physik C Particles and Fields **57** (1993) 485-494.

[20] K.K. Gudima, S.G. Mashnik, and A.J. Sierk, User Manual for the Code LAQGSM, LA-UR-01-6804 (2001) http://lib-www.lanl.gov/la-pubs/ 00818645.pdf.

[21] H. Sorge, H. Stocker, and W. Greiner, Nucl. Phys. **A 498** (1989) 567-576.

[22] N.V. Mokhov and C.C. James, The MARS Code System User's Guide, Fermilab-FN-1058-APC (2018), https://mars.fnal.gov/.

[23] N. Mokhov *et al.*, *Prog. Nucl. Sci. Technol.* **4** (2014) 496–501.

[24] I.L. Rakhno, N.V. Mokhov, and S.I. Striganov, Modeling Heavy Ion Ionization Loss in the MARS15 Code, Fermilab-Conf-05-019-AD (2005).

[25] D.C. Wilson, C.A. Wingate, J.C. Goldstein, R.P. Godwin and N.V. Mokhov, *IEEE Proc. Part. Accel. Conf.* (1993), 3090-3092.

[26] N.A. Tahir, F. Burkart, R. Schmidt, A. Shutov, and A.R. Piriz, *Nucl. Inst. and Methods in Physics Research* **B 427** (2018) 70-86.

[27] https://www.ansys.com.

[28] D.J. Cagliostro, D.A. Mandell, L.A. Schwalbe, T.F. Adams, and E.J. Chapyak, *Int. J. Impact Engng,* **10** (1990) 81-92.

[29] L. Lucy, *Astron. J.*, **48** (1977) 1013-1024.

[30] https://www.lstc.com/products/ls-dyna/.

[31] V.E. Fortov et al., *Nucl. Sci. Eng.* **123** (1996) 169.

[32] M.J. Norgett, M.T. Robinson, I.M. Torrens, *Nucl. Eng. Des.* **33** (1975) 50.

[33] N.V. Mokhov, I.L. Rakhno. And S.I. Striganov, Simulation and Verification of DPA in Materials, in *Proc. Workshop on Appl. High Intensity Proton Accel,* World Scientific (2010) 128-131.

[34] https://www.njoy21.io/NJOY2016/.

[35] N.V. Mokhov, DPA Modelling in MARS15, *NBI2017-RaDIATE Workshop*, IVIL, Tokai, Japan (2017).

[36] R.E. Stoller, *J. Nucl. Mat.* **276** (2000) 22.

[37] K. Nordlund, *Nature Communications* **9** (2018) 1084, https://www.nature.com/articles/s41467-018-03415-5#Bib1.

[38] S. Roesler and M. Silari, *Review of Particle Physics*, Phys. Rev. **D98**, 030001 (2018) 519-520.

[39] A. Ferrari, P.R. Sala, A. Fasso, and J. Ranft, FLUKA: A Multi-Particle Transport Code, CERN-2005-010 (2005).

[40] G. Battistoni *et al.*, *Ann. Nucl. Energy* **82** (2015) 10–18. http://dx.doi.org/10.1016/j.anucene.2014.11.007.

[41] FLUKA, http://www.fluka.org.

[42] S. Agostinelli *et al.*, *Nucl. Instr. Methods A* **506**(3) (2003) 250–303. http://dx.doi.org/10.1016/S0168-9002(03)01368-8.

[43] J. Allison *et al.*, *IEEE Trans. Nucl. Sci.* **53**(1) (2006) 270–278. http://dx.doi.org/10.1109/TNS.2006.869826.

[44] J. Allison *et al., Nucl. Instr. Methods A* **835** (2016) 186-225.

[45] Geant4, http://geant4.cern.ch.

[46] N.V. Mokhov, in *Proc. IV All-Union Conf. Charged Part. Accel.* **2** (1975) 222.

[47] N.V. Mokhov, MARS10 Code System, Fermilab-FN-509 (1989).

[48] MARS15(2018), https://mars.fnal.gov/.

[49] P. Aarnio, *Decay and Transmutation of Nuclides*, CMS-NOTE-1998/086, CERN (1998).

[50] R. Brun and F. Rademakers, ROOT - An Object-Oriented Data Analysis Framework, *Proc. AIHENP`96 Workshop*, Lausanne, Sep. 1996, Nucl. Instr. Meth. in Phys. Res. A **389** (1997) 81-86, http://root.cern.ch/.

[51] T. Roberts and D. Kaplan, G4beamline Simulation Program for Matter-Dominated Beamlines, in Proc. IEEE Part. Accel. Conf. (2007)3468-3470.

[52] MAD – Methodical Accelerator Design, http://cern.ch/madx/.
[53] N.V. Mokhov and I.S. Tropin, MARS15-Based System for Beam Loss and Collimation Studies, in *Proc. ICFA Mini-Workshop on Tracking for Collimation*, Geneva, Switzerland, 30 October 2015, edited by S. Redaelli, CERN Yellow Reports: Conference Proceedings, Vol. 2/2018, CERN-2018-011-CP (CERN, Geneva, 2018), pp. 57 – 72, https://doi.org/10.23732/CYRCP-2018-011.
[54] Long-Baseline Neutrino Facility (LBNF) https://lbnf.fnal.gov/.
[55] Deep Underground Neutrino Experiment (DUNE), https://www.dunescience.org/.
[56] T. Goorley *et al.*, *Nucl. Technol.* **180** (3) (2012) 298–315. http://dx.doi.org/10.13182/NT11-135.
[57] D. Pelowitz et al., LANL LA-CP-00745 (2014).
[58] Monte Carlo Code Group, A General Monte Carlo N-Particle (MCNP) Transport Code, http://mcnp.lanl.gov/.
[59] D. Georgobiani *et al*, in *Proc. SATIF-13 on Shielding Aspects of Accelerator, Target and Irradiation Facilities*, Dresden, (2016) 196-203.
[60] MCAM: CAD/Image-Based Modeling Program for Nuclear and Radiation System, http://www.fds.org.cn/en/software/mcam_1.asp
[61] VISED: Visual Editor for MCNP, http://www.mcnpvised.com/visualeditor/visualeditor.html.
[62] K. Niita *et al.*, *Radiat. Meas.* **41** (9–10) (2006) 1080–1090. http://dx.doi.org/10.1016/j.radmeas.2006.07.013.
[63] T. Sato, Y. Iwamoto, S. Hashimoto, T. Ogawa, T. Furuta, S. Abe, T. Kai, P.E. Tsai, N. Matsuda, H. Iwase, N. Shigyo, L. Sihver and K. Niita, Features of Particle and Heavy Ion Transport Code System PHITS Version 3.02, *J. Nucl. Sci. Technol.* **55**, 684-690 (2018).
[64] PHITS, Particle and Heavy Ion Transport Code System, http://phits.jaea.go.jp/.
[65] IAEA1437 SUPERMC, Super Monte Carlo Simulation Program for Nuclear and radiation Process, http://www.oecd-nea.org/tools/abstract/detail/iaea1437/.
[66] T. Oyama *et al*, Nucl. Instr. Meth. in Phys. Res. B **434** (2018) 29-36.
[67] I.L. Rakhno, J. Hylen, P. Kasper, N.V. Mokhov, M. Quinn, S.I. Striganov, and K. Vaziri, *Nucl. Inst. and Methods in Physics Research* **B 414** (2018) 4-10.
[68] B. Auchmann *et al.*, *Phys. Rev. ST Accel. Beams* **18** (2015) 061002.
[69] E. Skordis *et al.*, Proc. IPAC 2015, Richmond (2015) 2116.
[70] A.I. Drozhdin and N.V. Mokhov, STRUCT User's Reference Manual, SSCL-MAN-0034 (1994).
[71] M.A. Kostin, O.E. Krivosheev, N.V. Mokhov, and I.S. Tropin, An Improved MAD-MARS Beam Line Builder: User's Guide, Fermilab-FN-0738-rev (2004).
[72] L. Nevay *et al.*, Proc. IPAC 2014, Dresden (2014) 182.
[73] A. Mereghetti *et al.*, Proc. IPAC 2012, New Orleans (2012) 2687.
[74] F. Schmidt, SixTrack Version 4.2.16: single particle tracking code treating transverse motion with synchrotron oscillations in a symplectic manner, CERN-SL-94-56 (1994), http://sixtrack.web.cern.ch/SixTrack.
[75] A. Mereghetti *et al.*, Proc. IPAC 2013, Shanghai (2013) 2657.